\documentclass[a4paper]{jpconf}
\usepackage{graphicx}
\begin{document}
\title{Modified gravity and dark matter}

\author{Jose A. R. Cembranos}

\address{Departamento de F\'{\i}sica Te\'orica I, Universidad Complutense de Madrid, E-28040 Madrid, Spain.}

\ead{cembra@fis.ucm.es}

\begin{abstract}
The fundamental nature of Dark Matter (DM) has not been established.
Indeed, beyond its gravitational effects, DM remains
undetected by present experiments. In this situation, it is
reasonable to wonder if other alternatives can effectively explain
the observations usually associated with the existence of DM.
The modification of the gravitational interaction has been studied
in this context from many different approaches. However, the
large amount of different astrophysical evidences makes difficult
to think that modified gravity can account for all these observations.
On the other hand, if such a modification introduces new degrees of freedom,
they may work as DM candidates. We will summarize the phenomenology of these
gravitational dark matter candidates by analyzing minimal models.
\end{abstract}

\section{Introduction}

Several astrophysical observations conclude that the main amount of the present matter content of our Universe is in form of unknown particles that are not included in the Standard Model (SM) of particles and interactions. There are many well motivated candidates to account for this missing matter problem. However, the introduction of new degrees of freedom is not only well motivated within the gravitational sector, but absolutely necessary. Indeed, the non-renormalizability and non-unitarity of the Einstein-Hilbert action (EHA) demads its modification at high energies. In this contribution, we will discuss that this correction required the introduction of new states. These new fields will typically interact very weakly with SM fields and behave as dark matter (DM).

\section{$R^2$-gravity}

However, the ultraviolet (UV) completion of the gravitational interaction is an open question, and it is difficult to make general statements about its phenomenology. We will work in a minimal approach with the purpose of capturing the fundamental physics associated with this question. The first corrections to the EHA at high energies, may be provided by the inclusion of four-derivative terms in the metric preserving general covariance. In addition to the usual massless spin-two graviton, these terms support a massive spin-two and a massive scalar, with a total of six new degrees of freedom \cite{Stelle:1976gc, Stelle:1977ry}. Although, four-derivative gravity is renormalizable, the massive spin-two modes are ghost-like particles associated with the breaking of causality, new unitarity violations, and inadmissible instabilities \cite{Simon:1991bm}.

On the contrary, if the correction is determined by a term proportional to the square of the scalar curvature, the gravitational interaction
is phenomenologically viable. The so called $R^2$-gravity is free of ghosts although it does not improve the UV problems of Einstein gravity. However, it introduces a new scalar degree of freedom, which illustrates our idea in a minimal way. Indeed, its mass $m_0$ is related to the only new constant in the action (\ref{model}):
\begin{eqnarray}
\label{model}
S_G&=&\int\sqrt{g}\left\{-\Lambda^4
-\frac{M_{\rm{Pla}}^2}{2}R
\,+\frac{M_{\rm{Pla}}^2}{12\,m_0^2}R^2
\,+\,...\,\right\}
\\
&&
\;\;\;\;\;
\;\;\;\;\;\;\;\;\underbrace{\;\;}_{\rm{DE}}
\;\;\;\underbrace{\;\;\;\;\;\;\;\;\;\;\,}_{\rm{EHA}}
\;\;\;\;\underbrace{\;\;\;\;\;\;\;\;\;\;\;\;\;\;}_{\rm{DM}}\nonumber
\end{eqnarray}
with $M_{\rm{Pla}}\equiv (8\pi G_N)^{-1/2}\simeq 2.4 \times 10^{18}$ GeV,
$\Lambda\simeq 2.3\times 10^{-3}$ eV, and the dots mean general UV corrections
that must be present in the model to complete the high energy theory.
In \cite{Cembranos:2008gj}, is has been shown that just the Action (\ref{model}) can
describe the late time evolution of our Universe, since the first term is able account for the
dark energy (DE), whereas the third term can explain the dark matter (DM).

The Einstein's Equations (EEs) associated with $R^2$-gravity \cite{Starobinsky:1980te,Gottlober:1990um}
can be written as (following notation from \cite{Cembranos:2005fi}):
\begin{eqnarray}\label{equation1}
&&\left[1-\frac{1}{3\,m_0^2}\,R\right]R_{\mu\nu}-
\frac{1}{2}\left[R-\frac{1}{6\,m_0^2}\,R^2\right]g_{\mu\nu}\nonumber\\
&&
-\,\,\mathcal{I}_{\alpha\beta\mu\nu}
\nabla^\alpha\nabla^\beta\left[\frac{1}{3\,m_0^2}\,R\right]
=\frac{T_{\mu\nu}}{M_{\rm{Pla}}^2}\,.
\end{eqnarray}
Here $\mathcal{I}_{\alpha\beta\mu\nu}\equiv \left(g_{\alpha\beta}g_{\mu\nu}-g_{\alpha\mu}g_{\beta\nu}\right)$.
For $c_1\ll 1$, the metric $g_{\mu\nu}=[1+c_1 \sin(m_0 t)] \eta_{\mu\nu}$ is solution of the above equation.
Indeed, the energy stored in the mentioned oscillations have a cold DM at leading order. This massive mode is
an independent degree of freedom that eventually will cluster and produce a successful structure formation
if it is produced properly.

\section{Misalignment mechanism}

The thermal production is expected to be affected by higher order corrections to the Action (\ref{model}). At temperatures
$T \gg \Lambda_G\equiv\sqrt{M_{\rm{Pla}} m_0}$, the complete UV gravitational interaction will be fundamental to study the cosmology.
Nevertheless, there are other sources for the abundance of this field that can be analyzed from Eq. (1). In the same case that other bosonic degrees of freedom \cite{axions, Cembranos:2012kk}, this particle may be produced by the so called {\it misalignment mechanism}. In general, the initial value of the scalar mode ($\phi_1$) would not coincide with the minimum of its potential ($\phi=0$) for $H(T)\gg m_0$. Below the temperature $T_1$ for which $3H(T_1)\simeq m_0$, $\phi$  oscillates around this minimum. In such a case, the initial number density: $n_\phi \sim m_0\phi_1^2/2$
(with $\phi_1=\sqrt{\left\langle \phi(T_1)^2\right\rangle}$ ), evolves as dust or non-relativistic matter. Indeed, as such a number density  scales as the entropy density of radiation ($s=2 \pi^2 g_{s1} T_1^3/45$) for adiabatic expansions, we can estimate the abundance with the following expression:
\begin{eqnarray}
\Omega_{\phi}h^2&\simeq&
\frac{(n_\phi/s)(s_0/\gamma_{s1})}{\rho_{crit}}\,m_0\,.
\end{eqnarray}
Here $\rho_{crit}\simeq 1.0540\times10^{4}\,\rm{eV}\, \rm{cm}^{-3}$ is the critical
density, $s_0=2970\,\rm{cm}^{-3}$ is the present entropy density of radiation,
and $\gamma_{s1}$ is a factor that takes into account the increment of entropy in a comoving volume
since the onset of the oscillations. we can estimate $T_1$ by solving $m_0=3H_1(T_1)$.
In particular, for a radiation dominated universe at $T_1$ ($3H_1=\pi(g_{e\,1}/10)^{1/2}T_1^2/M_{\rm{Pla}}$):
\begin{eqnarray}
\label{T1}
T_1
\simeq 15.5\,\rm{TeV}
\left[
\frac{m_0}{1\, \rm{eV}}
\right]^{\frac12}
\left[
\frac{100}{g_{e\,1}}
\right]^\frac14\,.
\end{eqnarray}
It impolies an abundance of:
\begin{eqnarray}
\Omega_{\phi}h^2&\simeq&
0.86\,
\left[
\frac{m_0}{1\, \rm{eV}}
\right]^{\frac12}
\left[
\frac{\phi_1}{10^{12}\, \rm{Gev}}
\right]^{2}
\left[
\frac{100\, g_{e\,1}^3}{(\gamma_{s1} g_{s1})^4}
\right]^{\frac14},
\label{abundance}
\end{eqnarray}
where $g_{e\,1}$ and $g_{s1}$ are the effective energy and entropy
number of relativistic degrees of freedom at $T_1$ respectively.
Initial values for the scalar mode of the order of $\phi_1\sim 10^{12}\;\rm{GeV}$
can account for the total DM abundance (see Fig. \ref{gravDM}). These values are
consistent with our perturbative approach of the background metric
$||\Delta g_{\mu\nu}/\hat{g}_{\mu\nu}||\leq 10^{-6}$.

\section{Torsion-balance measurements}

On the other hand, the new scalar graviton mediates an attractive Yukawa force between two non-relativistic particles of masses $M_1$ and $M_2$:
\begin{eqnarray}
\label{Yukawa}
V_{ab}&=&-\alpha\frac{1}{8 \pi M_{\rm{Pla}}^2}\frac{M_1 M_2}{r} e^{- m_0\,r}\,,
\end{eqnarray}
where $\alpha=1/3$ \cite{Stelle:1977ry}. Torsion-balance experiments introduce the following lower bound on the mass of the new scalar mediator \cite{Kapner:2006si,Adelberger:2006dh}:
\begin{eqnarray}
\label{yukawa}
m_0\geq 2.7 \times 10^{-3} \rm{eV}
 \;\;\;\;\;\;\;\rm{at}
 \;\;\;\;95 \%\;\; \rm{c.l.}
\end{eqnarray}
This constraint does not depend on the misalignment or any other supposition about the relic abundance.

\section{Cosmic rays}

\begin{figure}[bt]
\begin{center}
\resizebox{8.0cm}{!} {\includegraphics{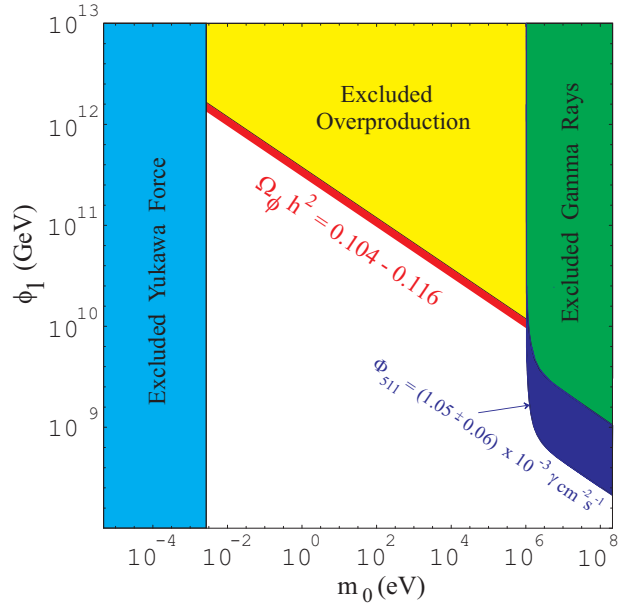}}
\caption{The parameter space of the model has two parameters: $m_0$ is the mass of the new scalar graviton, whereas
$\phi_1$ is its displacement with respect to the minimum of the potential for $3H\sim m_0$. In this figure,
$g_{e\,1}=g_{s1}\simeq 106.75$, and $\gamma_{s1}\simeq 1$ are assumed.
The left side is excluded by Torsion-balance experiments. The right one is ruled out by non-observation of cosmic-ray signals.
In the limit of this region, $R^2$-gravity may be the positron source to produce the 511 keV
line coming from the GC \cite{Knodlseder:2005yq}. The upper region is excluded by the overproduction of this new mode.
On the other hand, this mode can account for the total DM of the standard cosmological model on the diagonal line.
This figure has been taken from \cite{Cembranos:2008gj}.
}
\label{gravDM}
\end{center}
\end{figure}

Depending on such a relic abundance, $m_0$ may have an upper bound.
The $e^+e^-$ decay is the most constraining channel if $\phi$ constitutes the total DM.
Its decay rate can be written as:
\begin{eqnarray}
\label{positron}
\Gamma_{\phi\rightarrow e^+e^-}\simeq
\left[
2.14 \times 10^{24} s\; \frac{r_e^2}{\left(r_e^2-1\right)^{3/2}}
\right]^{-1}
\,,
\end{eqnarray}
where $r_e=m_0/(2 m_{e})$). The restriction is set by measurements of the SPI spectrometer on the INTEGRAL
(International Gamma-ray Astrophysics Laboratory) satellite. It has observed a 511 keV line emission
of $1.05 \pm 0.06 \times 10^{-3}$ photons cm$^{-2}$ s$^{-1}$ coming from the Galactic Center (GC)~\cite{Knodlseder:2005yq}.
This very monochromatic flux is consistent with the spectrum coming from $e^+ e^-$ annihilation and it agrees with previous
observations. However, the source of the positrons is unknown. Decaying DM could be a possible source for masses
$1\;\rm{MeV}\leq M_{\rm{DDM}}\leq 10\;\rm{MeV}$ \cite{Beacom:2005qv} if its decay rate in $e^+e^-$ verifies
\cite{Picciotto:2004rp, Hooper:2004qf, Kasuya:2006kj, Pospelov:2007xh, Cembranos:2008bw,Cembranos:2008gj}:
\begin{eqnarray}
\label{511}
\frac{\Omega_{\rm{DDM}}h^2\; \Gamma_{\rm{DDM}}}{M_{\rm{DDM}}} \simeq
\left[(0.2 - 4)\times 10^{27}\; \rm{s}\; \rm{MeV}\right]^{-1}
\,.
\end{eqnarray}
and the local DM structure has associated a cuspy profile \cite{Cembranos:2008bw}).
As one can see in Fig. \ref{gravDM}, $R^2$-gravity can explain this positron production, but
it demands a lower abundance unless if the mass $m_0$ is very close to $2\,m_e$.
If $m_0 \geq 1.2\; \rm{MeV}$, the new scalar graviton cannot constitute the total local DM since
it will originate a larger intensity for the 511 line. If $m_0 \geq 10\; \rm{MeV}$, the gamma rays
produced by inflight annihilation of the positrons with interstellar electrons sets more important
limits than the 511 keV flux \cite{Beacom:2005qv}. Finally, if $m_0 < 2\, m_e$, the only observable decay channel
is in two photons. The rate has been computed in \cite{Cembranos:2008gj}:
\begin{eqnarray}
\label{foton2}
\Gamma_{\phi\rightarrow \gamma\gamma}
\simeq \left[2.5 \times 10^{29} \rm{s}\left[\frac{1\, \rm{MeV}}{m_0}\right]^3\right]^{-1}
.\;\;\;\;
\end{eqnarray}

It is difficult to detect these photons in  the isotropic diffuse photon background (iDPB) if $m_0\leq 1\; \rm{MeV}$
\cite{Cembranos:2007fj,Cembranos:2008bw}. The analysis is associated with the search of gamma-ray lines
at $E_\gamma= m_0/2$ from localized sources is more promising, but only the heavier allowed region of the parameter space
may be explored with reasonable improvements of present detectors \cite{Cembranos:2007fj}.

\section{Conclusions}
\label{conclusions}

DM has been typically assumed to be in the form of stable Weakly-interacting massive
particles (WIMPs). They arise in well-motivated theoretical frameworks as in
supersymmetry (SUSY) models with R-parity conservation~\cite{SUSY1,SUSY2}, or
models with additional spatial dimensions, such as universal extra dimensions (UED)~\cite{UED1, UED2},
or brane-worlds~\cite{BW1,BW2,BW3}. However, there are many other possibilities.
In this work, we have analyzed the scenario in which the DM merges from high energy corrections of
the gravitational interaction.
We have illustrated this idea with the scalar mode associated with $R^2$-gravity,
but these type of fields are present in many different theories beyond the standard
model such as extra dimensions, supersymmety or string theory.
Another interesting feature of WIMPs, it is that they can be observed at the LHC or
in the new generation of colliders \cite{Coll}. This type of phenomenological options
are not promissing for this super-weakly interacting DM studied in this analysis. However,
Torsion-balance experiments or cosmic rays observations \cite{Cosmics} are sensitive to the gravitational
DM discussed in this analysis.

\section*
{Acknowledgments}
This work has been supported by MICINN (Spain) project numbers FIS2011-23000,
FPA2011-27853-01, FIS2014-52837-P, FPA2014-53375-C2-1-P and ConsoliderIngenio
MULTIDARK CSD2009-00064.

\end{document}